\documentclass[twocolumn]{autart}   
\usepackage{graphicx}         
\usepackage{url}
\usepackage{xcolor}
\usepackage{amsmath}
\usepackage{mathtools}
\usepackage{tikz}
\usepackage{verbatim}
\usepackage{subcaption}
\usepackage{pgfplots}
\usepackage{epstopdf}
\usepackage{amssymb}
\usepackage[textwidth=3.8em,textsize=scriptsize,disable]{todonotes}
\usepackage{algorithm}
\definecolor{Gray}{gray}{0.90}
\usepackage{colortbl}

\usepackage{algorithm}
\usepackage{algorithmicx}
\usepackage{algcompatible}
\usepackage{algpseudocode}
\usepackage{color}
\usepackage{blkarray}
\newtheorem{theorem}{\bf{Theorem}}[section]

\newtheorem{defi}[theorem]{\bf{Definition}}
\newtheorem{lemma}[theorem]{\bf{Lemma}}
\newtheorem{proposition}[theorem]{\bf{Proposition}}
\renewcommand{\qed}{\hfill\blacksquare}
\newtheorem{corollary}[theorem]{Corollary}
\newcommand{\calG}[0]{\mathcal{G}}
\newcommand{\calV}[0]{\mathcal{V}}
\newcommand{\calE}[0]{\mathcal{E}}

\newcommand{\calS}[0]{\mathcal{S}}

\newcommand{\calK}[0]{\mathcal{K}}

\newcommand\figref{Figure~\ref}

\definecolor{blue}{rgb}{0,0,0}
\definecolor{amber}{rgb}{1.0, 0.75, 0.0}
\begin{document}

\begin{frontmatter}
\title{Resilient Distributed Vector Consensus Using Centerpoint\thanksref{footnoteinfo}} 

\thanks[footnoteinfo]{Some preliminary results are presented in \cite{MudassirACC2020}.}

\author[a]{Waseem Abbas}\ead{waseem.abbas@utdallas.edu},
\author[b]{Mudassir Shabbir}\ead{mudassir.shabbir@vanderbilt.edu},
\author[b]{Jiani Li}\ead{jiani.li@vanderbilt.edu},              
\author[b]{Xenofon Koutsoukos}\ead{xenofon.koutsoukos@vanderbilt.edu}
\address[a]{Department of Systems Engineering, University of Texas at Dallas, Richardson, TX, USA.}   
\address[b]{Department of Electrical Engineering and Computer Science, Vanderbilt University, Nashville, TN, USA.}


\begin{abstract}                          
In this paper, we study the resilient vector consensus problem in networks with adversarial agents and improve resilience guarantees of existing algorithms. A common approach to achieving resilient vector consensus is that every non-adversarial (or normal) agent in the network updates its state 
by moving towards a point in the convex hull of its \emph{normal} neighbors' states. Since an agent cannot distinguish between its normal and adversarial neighbors, computing such a point, often called as \emph{safe point}, is a challenging task. 
To compute \textcolor{blue}{a safe point}, we propose to use the notion of \emph{centerpoint}, which is an extension of the median in higher dimensions, instead of Tverberg partition of points, which is often used for this purpose. We discuss that \textcolor{blue}{the notion of centerpoint provides a complete characterization of safe points in $\mathbb{R}^d$. In particular, we show that \textcolor{blue}{a safe point} is essentially an interior centerpoint if the number of adversaries in the neighborhood of a normal agent $i$ is less than $\frac{N_i}{d+1} $, where $d$ is the dimension of the state vector and $N_i$ is the total number of agents in the neighborhood of $i$.} Consequently, we obtain necessary and sufficient conditions on the number of adversarial agents to guarantee resilient vector consensus. Further, by considering the complexity of computing centerpoints, we discuss improvements in the resilience guarantees of vector consensus algorithms \textcolor{blue}{and compare with the other existing approaches}. Finally, we numerically evaluate the performance of our approach through experiments.

\end{abstract}

\end{frontmatter}

\section{Introduction}
\label{sec:Intro}
\begin{figure}
\centering
\includegraphics[scale=1.35]{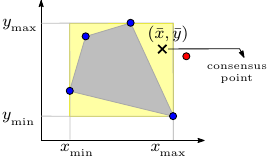}
\caption{Normal nodes (blue) implement resilient scalar consensus for $x$ and $y$ variables separately, and converge at $(\bar{x},\bar{y})$. Here, $\bar{x}\in [x_{\min}\;x_{\min}]$ and $\bar{y}\in [y_{\min}\;y_{\min}]$, as guaranteed by resilient scalar consensus. However, $(\bar{x},\bar{y})$ is outside the convex hull of normal nodes, which is a (gray) shaded region.}
\label{fig:rescon}
\end{figure}

Resilient consensus in a network of agents, some of which might be adversarial or faulty, has several applications in multirobot networks, distributed computing, estimation, learning and optimization (for instance, see \cite{abbas2017improving,li2018resilient,park2017fault,sundaram2015consensus,su2016multi,tseng2013iterative}). The main goal of resilient consensus is to ensure that all normal agents in a network agree on a common state despite the presence of some adversarial agents, whose identities are unknown to normal agents. Resilient consensus is achieved if appropriate state update laws are designed for normal agents and the underlying network topology satisfies certain connectivity and robustness conditions. For instance, when agents' states are scalars, \cite{leblanc2013resilient} presents a resilient distributed algorithm that guarantees convergence of normal agents to a common state $\bar{x}\in\left[x_{\min}\; x_{\max}\right]$, where $x_{\min}$ and $x_{\max}$ are the minimum and maximum of the initial values of normal agents, respectively. If agents' states are vectors or points in $\mathbb{R}^d$, $d\ge 2$, then the resilient consensus objective is to ensure that normal agents converge at some point in the convex hull of their initial states. \textcolor{blue}{One such application is the resilient multirobot rendezvous problem, where the goal is to gather mobile robots dispersed in a region at a common point in a distributed manner \cite{park2017fault,mendes2015multidimensional}. The robots must move within their original convex region to ensure that they do not drift to an unsafe region.} A simple approach could be to run $d$ instances of scalar resilient consensus, one for each dimension. However, as a result of this approach, normal agents might converge at a point outside the convex hull of their initial states, as discussed in \cite{Vaidya:2013:BVC:2484239.2484256}. We also illustrate such a situation in Figure \ref{fig:rescon}. As a result, we cannot rely on resilient scalar consensus algorithms to achieve resilient vector consensus.

In this paper, we study the resilient vector consensus problem and propose a \emph{resilient aggregation rule} for normal (non-adversarial) agents to update their states by combining their neighbors' states, which might include adversarial agents. In the proposed resilient aggregation rule, each normal agent computes a \emph{centerpoint},  which is an extension of the median in the higher dimensional Euclidean space, of its neighbors' states. We then obtain necessary and sufficient conditions on the number of adversarial agents each normal node can have in its neighborhood, while guaranteeing resilient vector consensus. We show that the centerpoint-based aggregation improves the resilience of the known vector consensus algorithms, for instance, the \emph{Approximate Distributed Robust Convergence (ADRC)} algorithm recently proposed in \cite{park2017fault}. We note that centerpoint of points in $\mathbb{R}^d$ is different from taking the coordinate-wise median of points.

Various resilient vector consensus algorithms have been proposed in the literature, for instance, see \cite{park2017fault,Vaidya:2013:BVC:2484239.2484256,vaidya2014iterative,mendes2015multidimensional,park2013worst,yan2020resilient,yan2019safe}. The main idea in these algorithms is that each normal agent updates its state iteratively by computing a state (point in $\mathbb{R}^d$) that is guaranteed to lie in the \emph{interior of the convex hull of its normal neighbors'} states. Since a normal agent cannot distinguish between its normal and adversarial neighbors, computing such a point, which is often referred to as the \emph{safe point}, is the primary challenge in these consensus algorithms. To compute a safe point, previous works \cite{park2017fault,mendes2015multidimensional} have utilized the idea of \emph{Tverberg partition} of points in $\mathbb{R}^d$ (discussed in Section \ref{sec:ADRC}). We argue that instead of computing Tverberg partition, it is better to use the notion of \emph{centerpoint} in $\mathbb{R}^d$ to compute safe points. We show that safe point is essentially an interior centerpoint of the neighbors' states. This perspective provides a complete characterization of safe points, and hence allows us to improve the resilience bound of the vector consensus algorithms. A centerpoint has been known to the discrete geometry community for a long time, and \textcolor{blue}{its} properties and generalizations are still an active topic of research \cite{shabbir2014some,mustafa2015k}. We summarize our contributions below:

\begin{itemize}
\item For a given set of $N_i$ points in $\mathbb{R}^d$, of which any $F_i$ are adversarial,  \textcolor{blue}{we show the equivalence between the notion of a safe point and an interior centerpoint for$N_i > F_i(d+1) $}. Here, $F_i$ and ${N}_i$ denote the number of adversarial agents and the total number of agents in the neighborhood of a normal agent $i$, respectively. Using this relationship, we discuss how the resilience of the vector consensus algorithm can be improved using centerpoint to compute a safe point.

\item We generalize a sufficient condition for the existence of safe point
to arbitrary $d>0$, where $d$ is the dimension of the state vector. \textcolor{blue}{In particular, we show that if $F_i < \lfloor \frac{{N}_i}{d+1}\rfloor$, then there always exists an \emph{interior centerpoint}, and hence, a safe point. A similar condition for the existence of safe point was established in \cite{park2017fault} for $d \le 8$ using the idea of Tverberg partition.
}
\textcolor{blue}{
\item Using the notion of centerpoint, we show that ${N}_i > (d+1)F_i$ is not only sufficient but also a necessary condition for computing a safe point.
}
\textcolor{blue}{
\item We show that the centerpoint-based vector consensus algorithm is resilient to  $F_i < \lfloor \frac{{N}_i}{d+1}\rfloor$ adversaries in the neighborhood of a normal agent $i$ in $d=2,3$. In higher dimensions ($d>3$), the centerpoint-based algorithm is resilient to $F_i$ adversaries, where $F_i < F_{max} = \Omega(\frac{{N}_i}{d^2})$. This approach offers improved resilience compared to the Tverberg partition-based algorithm, where $F_{max} = \Omega(\frac{{N}_i}{2^d})$.
}
\item We numerically evaluate and compare our results with other algorithms by simulating resilient vector consensus in multirobot networks.
\end{itemize}

In the case of resilient \emph{scalar} consensus, the \emph{trimming} method has been an effective approach to aggregate neighbors' states. The basic idea in this scheme is that a normal agent collects its neighbors' states (scalar), sorts them, and then disregards \emph{extreme} values, that is, $F$ largest and smallest values, where $F$ is an upper bound on the maximum number of adversarial agents in the neighborhood of a normal agent. Depending on the problem setup and threat model, $F$ can also be an upper bound on the total number of adversarial agents in the network. The normal agent then updates its state by aggregating states of the remaining neighbors, for instance, by computing their average or median. This approach has been applied to achieve resilient consensus in a wide variety of setups, attack scenarios, and network conditions in various application domains including distributed computing \cite{dolev1986reaching,kieckhafer1993low,kieckhafer1994reaching}, multirobot systems \cite{bouzid2009byzantine,bouzid2010optimal,saldana2019resilient,saulnier2017resilient}, network control \cite{abbas2017improving,leblanc2013resilient,dibaji2015consensus,dibaji2017resilient,saldana2017resilient,usevitch2019resilient,zhang2012simple}, and distributed optimization \cite{li2018resilient,furesilient,ravi2019case,sundaram2016secure,sundaram2018distributed,yang2019byrdie}. The trimming approach, however, is not directly applicable to the resilient vector consensus as the appropriate notion of vector states that are `farthest' from a normal agent's own state is difficult to define. Sorting states by simply computing the Euclidean distance between them in $\mathbb{R}^d$ does not work and more advanced ways need to be defined for this purpose \cite{Kuwaranancharoen2020,WangNLA2019}. At the same time, implementing resilient scalar consensus using the trimming approach in each dimension separately does not always solve the resilient vector consensus problem as we discuss earlier (Figure \ref{fig:rescon}). Consequently, in the resilient vector consensus case, the notion of \emph{safe point}, which is a point obtained by a convex combination of node's normal neighbors' states, is quite useful. Using the idea of safe point, various algorithms and resilience bounds have been proposed in the literature for the resilient vector consensus and distributed optimization problems \cite{tseng2013iterative,Vaidya:2013:BVC:2484239.2484256,vaidya2014iterative,mendes2015multidimensional,park2013worst,yan2020resilient,yan2019safe,LI-RSS-20}. Here, we focus on computational aspects and geometric characterization of safe points.

The rest of the paper is organized as follows: Section \ref{sec:Notations} introduces notations and preliminaries. Section \ref{sec:ADRC} provides an overview of the resilient vector consensus algorithms, in particular, the ADRC algorithm and its resilience bounds. Section \ref{sec:ADRC_using_CP} discusses the notion of centerpoint for resilient vector consensus and presents main results in the paper. Section \ref{sec:Evaluation} presents a numerical evaluation of our results, and Section \ref{sec:Con} concludes the paper.

\section{Notations and Preliminaries}
\label{sec:Notations}
We consider a network of agents modeled by a \emph{directed graph} $\mathcal{G}=(\mathcal{V},\mathcal{E})$, where $\mathcal{V}$ represents agents and $\mathcal{E}$ represents interactions between agents. Each agent $i\in\mathcal{V}$ has a $d$-dimensional state vector whose value is updated over time. The state of agent $i$ {at time $t$} is represented by {a point $x_i(t) \in \mathbb{R}^d$}. An edge $(j,i)$ means that $i$ can observe the state of $j$. The \emph{neighborhood} of $i$ is the set of nodes $\mathcal{N}_i =~\{j\in~\mathcal{V}| (j,i)\in\mathcal{E}\}\cup\{i\}$. 
For a given set of points ${X}\subset\mathbb{R}^d$, we denote its \emph{convex hull} by conv$({X})$. {A set of points in $\mathbb{R}^d$ is said to be in \emph{general positions} if no hyperplane of dimension $d-1$ or less contains more than $d$ points}. A point $x\in\mathbb{R}^d$ is an \emph{interior point} of a set ${X}\subset\mathbb{R}^d$ if there exists an open ball centered at $x$ which is completely contained in ${X}$. We use terms \emph{agents} and \emph{nodes} interchangeably, and similarly use terms \emph{points} and \emph{states} interchangeably.   

\emph{Normal and Adversarial Agents --} There are two types of agents in the network, {normal} and {adversarial}. \emph{Normal} agents, denoted by $\mathcal{V}_n\subseteq \mathcal{V}$, are the ones that interact with their neighbors synchronously and always update their states according to a pre-defined state update rule, which is the consensus algorithm. \emph{Adversarial} agents, denoted by $\mathcal{V}_f\subset \mathcal{V}$, are the ones that can change their states arbitrarily and do not follow the pre-defined state update rule. Moreover, an adversarial node can transmit different values to its different neighbors, which is referred to as the \emph{Byzantine} model. The number of adversarial nodes in the neighborhood of a normal node $i$ is denoted by $F_i$. For a normal node $i$, all nodes in its neighborhood are indistinguishable, that is, $i$ cannot identify which of its neighbors are adversarial.

\emph{Resilient Vector Consensus --} The goal of the resilient vector consensus is to ensure the following conditions:
\begin{itemize}
\item \emph{Safety --} Let ${X}(0) = \{x_1(0),x_2(0),\cdots,x_n(0)\}\subset\mathbb{R}^d$ be the set of initial states of normal nodes, then at each time step $t$, and for any normal node $i$, the state value of $i$, denoted by $x_i(t)$ should be in the conv$({X}(0))$. 
\item \emph{Agreement --} For every $\epsilon > 0$, there exists some $t_\epsilon$, such that for any normal node pair $i,j$, $||x_i(t) -~ x_j(t)|| < ~\epsilon$, $\forall t > t_\epsilon$.
\end{itemize}

\section{Background and Resilient Vector Consensus Algorithm} 
\label{sec:ADRC}
In this section, we provide an overview of resilient vector consensus algorithms in which the main idea is that each normal agent updates its state by moving towards a point in the convex hull of its normal neighbors. In particular, we discuss the resilient vector consensus algorithm, known as the \emph{approximate distributed robust convergence (ADRC)}, recently proposed in \cite{park2017fault}. Then, we discuss improvement in the resilience guarantees of the algorithm by reconsidering its computational aspects. \textcolor{blue}{We note here that our results are general and mainly provide an improved way for normal agents to aggregate their neighbors' states in the presence of adversarial agents. Consequently, these results can be applied to a broader class of algorithms that require each normal agent in the network to aggregate its neighbors' states, such that the aggregated state is guaranteed to lie in the convex hull of its normal neighbors, for instance, as in resilient distributed diffusion \cite{LI-RSS-20,sayed2013diffusion}}.

The ADRC algorithm is a synchronous iterative algorithm that guarantees the consensus of normal agents if the number of adversarial agents in the neighborhood of each normal agent is bounded by a certain value that depends on $d$. The notion of \emph{$F$-safe point} is crucial to understanding the algorithm.

\begin{defi} (\emph{$F$-safe point})
\label{def:safe}
Given a set of $N$ points in $\mathbb{R}^d$, of which at most $F$ are adversarial, then a point $s$ that is guaranteed to lie in the interior of the convex hull of $(N-F)$ normal points is an $F$-safe point.
\end{defi}

The ADRC algorithm relies on the computation of an $F_i(t)$-safe point in each iteration $t$ by every normal agent $i$ having $N_i(t)$ agents in its neighborhood, of which at most $F_i(t)$ are adversaries.\footnote{For simplicity, when the context is clear, we use $N_i$ and $F_i$ instead of $N_i(t)$ and $F_i(t)$, respectively.} Each normal agent $i$ updates its state as follows \cite{park2017fault}:

\begin{itemize}
    \item In the iteration $t$, a normal agent $i$ gathers the state values of its neighbors, and then computes an $F_i(t)$-safe point, denoted by $s_i$.
    \item Agent $i$ then updates \textcolor{blue}{its} state as below.
    \begin{equation}
        \label{eq:ADRC}
        x_i(t+1) = \alpha_i(t)s_i(t) + (1 - \alpha_i(t))x_i(t),
    \end{equation}
    where $\alpha_i(t)$ is a dynamically chosen parameter in the range (0  1), whose value depends on the application~\cite{park2017fault}.
\end{itemize}

\textcolor{blue}{\emph{Convergence --} It is shown in \cite{park2017fault} that if all normal agents update their states according to \eqref{eq:ADRC} after computing \emph{$F_i(t)$-safe points} in each iteration $t$ and the sequence of graphs induced by normal agents is \emph{repeatedly reachable} (as defined below)}, then all normal agents converge at a common point and achieve resilient consensus. \textcolor{blue}{The notion of repeatedly reachable sequence of graphs is defined in \cite{park2017fault}. We state it here also for completeness.}

{\color{blue} Let $\bar{\mathcal{V}}$ be the set of \emph{normal} agents in the network and $\bar{\mathcal{G}}(t) = (\bar{\mathcal{V}},\mathcal{E}(t))$ be the graph induced by normal agents at time $t$.

\begin{defi}{(Jointly Reachable Sequence of Graphs)}
\label{def:jointly_sequence_graphs}
Let $j$ be a non-negative integer. A finite sequence of graphs $\bar{\calG}(T_j), \bar{\calG}(T_j+1) \cdots, \bar{\calG}(T_{j+1}-1) $, where each graph in the sequence has the same vertex set $\bar{\calV}$ is called jointly reachable if the union of graphs defined as
$$
\bigcup\limits_{t=T_j}^{T_{j+1}-1}\bar{\calG}(t) = \left(\bar{\calV},\bigcup\limits_{t=T_j}^{T_{j+1}-1}\bar{\calE}(t)\right)
$$
contains a vertex $v\in\bar{\calV}$ such that for every $v'\ne v$ there exists a path form $v'$ to $v$ in this union of graphs.
\end{defi}

\begin{defi}{(Repeatedly Reachable Sequence of Graphs)}
\label{def:repeatedly_sequence_graphs}
An infinite sequence of graphs $\bar{\calG}(0), \bar{\calG}(1) \cdots$ is called repeatedly reachable if there is a sequence of times $0=T_1 < T_2 < T_3 \cdots$ such that $T_{j+1} - T_j < \infty$ and the subsequence $\bar{\calG}(T_j), \bar{\calG}(T_j + 1),\cdots \bar{\calG}(T_{j+1} - 1)$ is jointly reachable for all $j$.
\end{defi}

In simple words, an infinite sequence $\bar{\calG}(0), \bar{\calG}(1) \cdots$ is repeatedly reachable if it can be partitioned into contiguous finite length subsequences that are themselves jointly reachable.

We note that two types of conditions need to be satisfied for resilient consensus: (a) The underlying network topology should be such that the sequence of graphs induced by normal agents is repeatedly reachable. (b) Each normal agent can compute a safe point in each iteration. Our focus in this paper is on the computation of a safe point, including conditions under which it is possible.
}

\begin{figure}[h]
\centering
\includegraphics[scale=0.58]{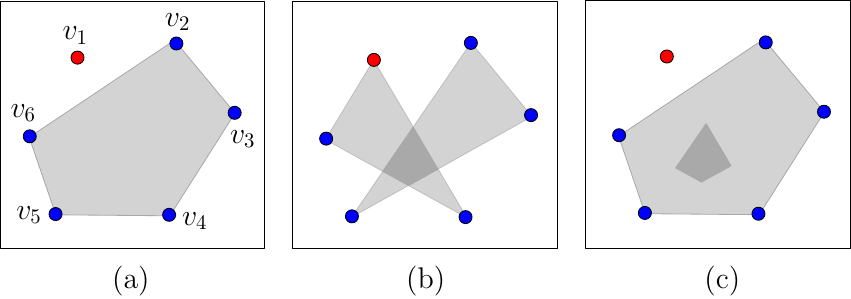}
\caption{{(a) Five normal (blue) and a single adversarial (red) node. Shaded area is the convex hull of normal nodes. (b) A Tverberg partition consisting of two subsets, of which one contains only normal nodes. Convex hulls of both subsets have a non-empty intersection, corresponding to a Tverberg region (dark shaded). (c) The intersection of Tverberg region and the convex hull of normal nodes.}}
\label{fig:intro}
\end{figure}

\textcolor{blue}{\emph{Computation of Safe point --}} It is the key step in the ADRC algorithm. For this, \cite{park2017fault} utilizes the notion of \emph{Tverberg partition} \cite{tverberg1966generalization}. The main idea is to partition a set of $N$ points in $\mathbb{R}^d$ into $(F+1)$ subsets such that the convex hull of points in one subset has a non-empty intersection with the convex hull of points in any other subset. The intersection of convex hulls of these $(F+1)$ subsests is a \emph{Tverberg region} and any interior point in such a region is an $F$-safe point, as illustrated in Figure \ref{fig:intro}. To see this, observe that if there are at most $F$ points corresponding to adversarial agents, then there is at least one subset in the partition that consists of points corresponding to normal agents only. The convex hull of points in that subset and therefore, the Tverberg region is guaranteed to lie in the convex hull of points corresponding to normal agents, which means that an interior point in the Tverberg region is an $F$-safe point.

Thus, to achieve resilient consensus using ADRC algorithm, a normal agent $i$ 
computes an $F_i$-safe point by computing an interior point in the Tverberg region of an appropriate Tverberg partition. Now, for given $N_i$, $F_i$ and $d$, the condition guaranteeing the existence of Tverberg partition is stated below \cite{park2017fault,reay1968,roudneff1990partitions,roudneff2009new}.

\begin{theorem} (Sufficient condition for $F_i$-safe point)
\label{thm:Park_Suff}
Given a set of $N_i$ points in general positions in $\mathbb{R}^d$ where $d\in\{2,3,\cdots,8\}$, of which at most $F_i$ points correspond to adversarial nodes, then it is possible to find an $F_i$-safe point if 
\begin{equation}
\label{eqn:theory}
F_i \le \frac{N_i}{d+1}- 1.
\end{equation}
\end{theorem}

\textcolor{blue}{In general, computing a Tverberg partition is an NP-hard problem and it could take $d^{O(d^2)}N_i^{(d+1)^2+1}$ time to compute a Tverberg partition of $N_i$ points in $\mathbb{R}^d$ into $\lceil N_i/(d+1)\rceil$ parts \cite{har2020improved}. The best known algorithm that computes it in a reasonable run time is an approximate algorithm \cite{mulzer2013approximating}, which has a time complexity of $d^{O(1)}N_i$.} The algorithm is approximate in a sense that to have a partition of $N_i$ points into $r$ subsets, $N_i\ge 2^d r$ (compared to the theoretical bound, $N_i\ge r(d+1)$). Consequently, to compute an $F_i$-safe point in the presence of $F_i$ adversarial neighbors, a normal agent $i$ needs to have at least $N_i \ge (F_i +1)2^d$ agents in its neighborhood, which gives the following resilience bound \cite{park2017fault}.

\begin{theorem} (Practical resilience bound)
\label{thm:Tvb_practical}
Given a set of $N_i$ points in general positions in $\mathbb{R}^d$, of which at most $F_i$ are adversarial, then it is possible to compute an $F_i$-safe point using Tverberg partition if
\label{prop:practical}
\begin{equation}
    \label{eq:practical}
    F_i \le \left\lceil\frac{N_i}{2^d}\right\rceil - 1.
\end{equation}
\end{theorem}

Note that \eqref{eq:practical} indicates \emph{resilience} of the ADRC algorithm based on approximate Tverberg partitions to compute safe points. For instance, the algorithm guarantees resilient consensus in $\mathbb{R}^2$ if for every normal agent, less than $25\%$ of its neighbors are adversarial.

Here, we ask if it is possible to improve the resilience of the ADRC algorithm. In particular, we study the following problems in the rest of the paper.
\begin{enumerate}
\item ADRC algorithm based on approximate Tverberg partition is resilient against $\Omega\left(\frac{N_i}{2^d}\right)$ adversaries in the neighborhood of a normal agent $i$. Can we improve the resilience by computing safe points some other way?

\item We have a sufficient condition for an $F_i$-safe point in \eqref{eqn:theory}. Can we provide a complete characterization of an $F_i$-safe point? In other words, what is a necessary and sufficient condition for a point to be an $F_i$-safe point?

\item The sufficient condition in \eqref{eqn:theory} holds for $d\le 8$. Can we generalize results to dimensions $d > 8$?
\end{enumerate}

\textcolor{blue}{The primary difference between our proposed approach and previous methods is the characterization and computation of safe points. There are other algorithms in the literature to compute safe points, for instance, \cite{park2017fault,mendes2015multidimensional,yan2020resilient,WangNLA2019}. We discuss in detail that the centerpoint-based approach is better for the characterization and computation of safe points and also compare this approach with the existing methods in Section \ref{sec:new_compare}.  }

\section{Resilient Vector Consensus Using Centerpoint}
\label{sec:ADRC_using_CP}
In this section, we show that it is possible to improve the resilience of ADRC algorithm by computing safe points differently, that is, by using the notion of \emph{centerpoint} instead of Tverberg partition. Moreover, centerpoint provides a better characterization of necessary and sufficient conditions for computing safe points.

\subsection{$F$-Safe Point and Interior Centerpoint}
The notion of centerpoint can be viewed as an extension of median in higher dimensions and is defined below.

\begin{defi}  (Centerpoint)
Given a set $\calS$ of $N$ points in $\mathbb{R}^d$ in general positions, a centerpoint $p$ is a point, such that any closed halfspace\footnote{Recall that closed halfspace in $\mathbb{R}^d$ is a set of the form $\{x\in\mathbb{R}^d: a^Tx \ge b\}$ for some $a\in \mathbb{R}^d\setminus \{0\}$.} of $\mathbb{R}^d$ that contains $p$ also contains at least $\frac{N}{d+1}$ points from $\calS$. Note that centerpoint does not need to be one of the points in $\calS$.
\end{defi}

\begin{figure}[h]
\centering
\begin{subfigure}[b]{0.22\textwidth}
\centering
\includegraphics[scale=0.46]{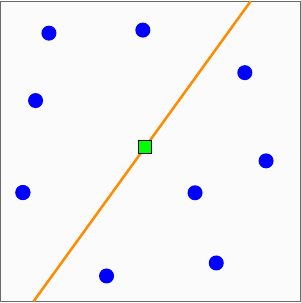}
\caption{$d=2$}
\end{subfigure}
\begin{subfigure}[b]{0.22\textwidth}
\centering
\includegraphics[scale=0.44]{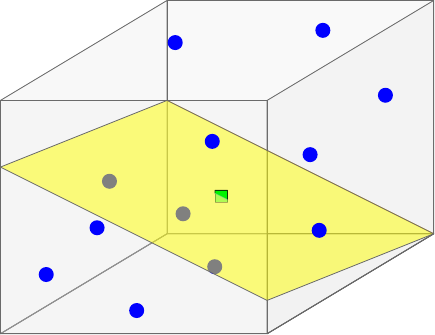}
\caption{$d=3$}
\end{subfigure}
\caption{Illustration of centerpoints (green squares).}
\label{fig:CP_example}
\end{figure}

Intuitively, a centerpoint lies in the `center region' of the point cloud, in the sense that  there are enough points of $\calS$ on each side of a centerpoint. More precisely, it is a point such that any hyperplane that goes through this point divides the set of points in $\mathcal{S}$ into two subsets, each of which contains at least $\frac{1}{d+1}$ fraction of points.  Figure~\ref{fig:CP_example} illustrates centerpoint in $\mathbb{R}^2$ and $\mathbb{R}^3$. In Figure~\ref{fig:CP_example}(a), any line through a centerpoint divides a set of 9 points into two subsets, each of which contains at least 3 points. Similarly, any plane through a centerpoint in Figure~\ref{fig:CP_example}(b) divides a set of 12 points into two parts, each containing at least 3 points.

A related notion of \emph{point depth} is defined as follows:

\begin{defi}  (Point depth)
For a given point set, the \textit{depth} of a point $q$ is the maximum number $\alpha$ such that every closed halfspace containing $q$ contains at least $\alpha$ points.
\end{defi}

Thus, a centerpoint has depth at least $\frac{N}{d+1}$. A centerpoint does not need to be unique; in fact, there can be infinitely many centerpoints. The set of all centerpoints constitutes the \textit{centerpoint region} or simply the \textit{center region}, which is closed and convex. The existence of a center point for any given set $\calS$ is guaranteed by the famous Centerpoint Theorem, which asserts that for any given point set in general positions in an arbitrary dimension, a centerpoint always exists. (see \cite{matouvsek2002lectures,rado1946theorem}).

We show that there is a close relationship between an $F$-safe point and an interior centerpoint, which can be exploited to provide a complete characterization of $F$-safe points. We state the following result:

\begin{theorem}
\label{thm:safe_center}
\textcolor{blue}{
For a given set of $N_i$ points in $\mathbb{R}^d$ (corresponding to $N_i$ agents in the neighborhood of agent $i$), if $N_i > F_i(d+1)$, then an $F_i$-safe point is equivalent to an interior centerpoint.
}
\end{theorem}

\emph{Proof:} For the proof of this statement, we observe that every centerpoint of a given set of $N_i$ points lies in the intersection of all convex sets that contain more than $\frac{N_id}{d+1}$ points \cite{matouvsek2002lectures}. To see this, imagine a convex set $C$ that does not contain a centerpoint $p$. Since $C$ is convex and $p$ is an outside point, there exists a hyperplane $H$ that separates $C$ from $p$. The hyperplane $H$ defines two halfspaces of $\mathbb{R}^d$: one halfspace contains $p$, and the other contains $C$. The halfspace that contains $C$ contains more than $\frac{N_id}{d+1}$ of the given points. Therefore, the halfspace that contains $p$ must contain less than $\frac{N_i}{d+1}$ of the given points, which is a contradiction because $p$ is a centerpoint. It implies that every convex set containing more than $\frac{N_id}{d+1}$ points must contain every centerpoint. Thus, all centerpoints lie in the intersection of such convex sets.

Now, we consider a agent $i$ having $N_i$ agents in its neighborhood. 
\textcolor{blue}{ An $\left(\frac{N_i-1}{d+1}\right)$-safe point lies in the interior of convex hull of points corresponding to normal agents when there are at most $\frac{N_i-1}{d+1}$ adversarial agents in the neighborhood of agent $i$.} As there are more than $\frac{N_id}{d+1}$ normal neighbors of $i$, an interior centerpoint lies in the relative interior of convex hull of normal agents' points. Therefore, every interior centerpoint is an $F_i$-safe point.
\textcolor{blue}{
On the other hand, if there are $\frac{N_i-1}{d+1}$ points corresponding to adversarial agents (whose identities are unknown) in the neighborhood of $i$; then an $F_i$-safe point, where $F_i=\frac{N_i-1}{d+1}$, must lie in the interior of an arbitrary convex set {$X$} containing more than $\frac{N_id}{d+1}$ points.}{This is true because otherwise an adversary can choose all agents corresponding to points in $X$ to be normal and all remaining agents to be adversarial to leave a chosen point `unsafe'.} It follows that every $F_i$-safe point must lie in the interior of the intersection of convex sets that contain more than $\frac{N_id}{d+1}$ points. Thus, every $F_i$-safe point must be an interior centerpoint. $\qed$

\begin{figure}[htb]
\centering
\begin{subfigure}[b]{0.115\textwidth}
\centering
\includegraphics[scale=0.417]{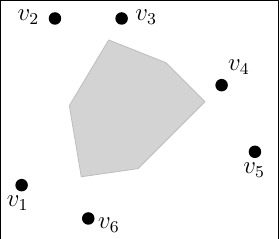}
\caption{}
\end{subfigure}
\begin{subfigure}[b]{0.115\textwidth}
\centering
\includegraphics[scale=0.417]{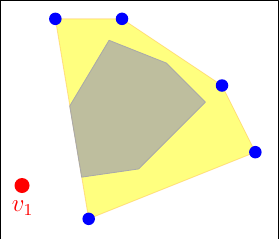}
\caption{}
\end{subfigure}
\begin{subfigure}[b]{0.115\textwidth}
\centering
\includegraphics[scale=0.417]{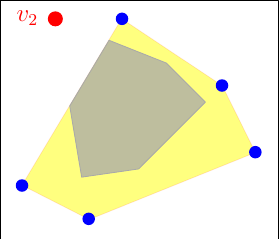}
\caption{}
\end{subfigure}
\begin{subfigure}[b]{0.115\textwidth}
\centering
\includegraphics[scale=0.417]{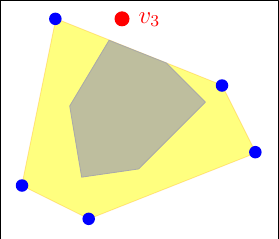}
\caption{}
\end{subfigure}
\begin{subfigure}[b]{0.125\textwidth}
\centering
\includegraphics[scale=0.417]{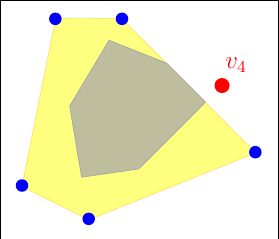}
\caption{}
\end{subfigure}
\begin{subfigure}[b]{0.125\textwidth}
\centering
\includegraphics[scale=0.417]{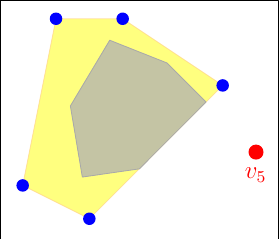}
\caption{}
\end{subfigure}
\begin{subfigure}[b]{0.125\textwidth}
\centering
\includegraphics[scale=0.417]{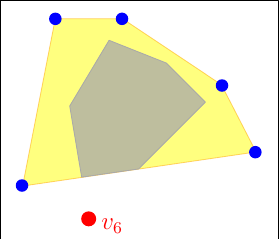}
\caption{}
\end{subfigure}
\caption{The region of centerpoints (gray area) is same as the region of $1$-safe points.}
\label{fig:CE}
\end{figure}

We illustrate the equivalence between $F_i$-safe point and centerpoint through an example. Consider six points in $\mathbb{R}^2$ in Figure \ref{fig:CE}(a). The gray region is the set of all centerpoints. At the same time, the gray region is also the set of all $1$-safe points. It means that no matter which one of the six points corresponds to an adversarial agent, every point in the the gray region is guaranteed to lie in the convex hull of remaining five points corresponding to normal agents. For example, in Figure \ref{fig:CE}(b), the red point is an adversary and the yellow region is the convex hull of normal agents' points. We observe that all $1$-safe points (gray region) lie inside the convex hull of normal agents' points. The same is illustrated for other possibilities for a single adversary point in Figures~\ref{fig:CE}(c)--(g).

\subsection{Necessary Condition for $F_i$-Safe Point}
\begin{figure}[htb]
	\centering
	\begin{subfigure}[b]{0.236\textwidth}
		\centering
		\includegraphics[scale=0.3]{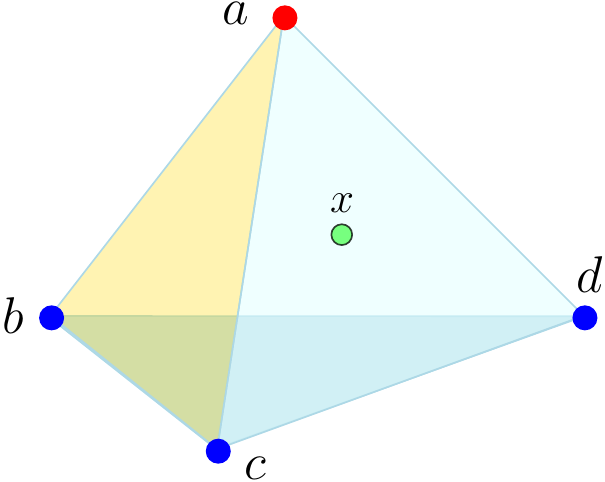}
		\caption{}
	\end{subfigure}
	\begin{subfigure}[b]{0.236\textwidth}
		\centering
		\includegraphics[scale=0.3]{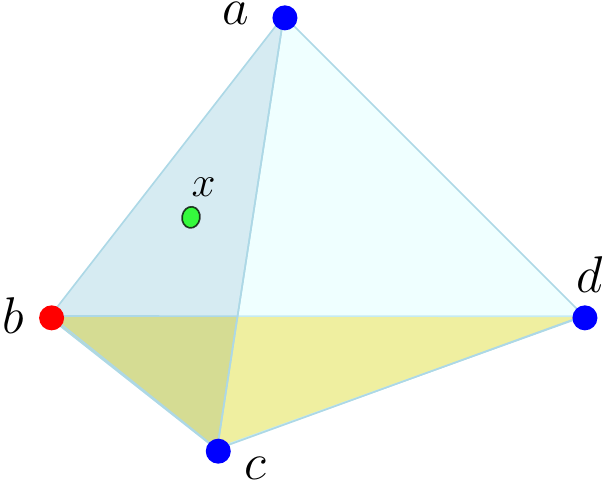}
		\caption{}
	\end{subfigure}
	\caption{An example of $4$ points in $\mathbb{R}^3$. (a)~A point $x$ in the interior of tetrahedron can not be $1$-safe because if $a$ is adversarial, then $x$ is not contained in the convex hull of the normal points. (b)~If a point $x$ lies on facet, say $a,b,c$, and $b$ is adversarial, then $x$ is not contained in the convex hull of normal point $a,c,d$. }
	\label{fig:NC}
\end{figure}

We note that $\frac{1}{d+1} $ is the best possible fraction (for $F_i$) in Theorem \ref{thm:safe_center}, that is, there exist general node positions where allowing more adversary nodes would mean that there is no safe point at all. For example, consider a set of three points $a,b,c,$ in $\mathbb{R}^2$ that lie on the vertices of a non-degenerate triangle.\footnote{A non-degenerate triangle has three points that do not all lie on a line.} If one of these points is malicious then there is no $1$-safe point in the plane. Clearly, no point in the interior of the triangle can be $1$-safe because it does not lie in the convex hull of normal points for any choice of point corresponding to an adversary. Similarly, any point that lies on an edge of the triangle, say $\overline{ab}$, can not be $1$-safe because if one of the endpoint, say $a$, is adversarial then the convex hull of the remaining points $b$ and $c$ does not contain this point. Now in $\mathbb{R}^3$, consider four points $a,b,c,d,$ lying on the vertices of a non-degenerate tetrahedron. Assume one of these points, whose identity is unknown, is adversarial. {A point $x$ in the interior of tetrahedron} can not be $1$-safe because if $a$ is malicious then {$x$} is not contained in the convex hull of normal points $b, c ,d$, as shown in Figure~\ref{fig:NC}(a). Similarly, consider a point $x$ that lies on a facet, say $a,b,c$, of the tetrahedron. The point $x$ can not be $1$-safe because if one of the points on \textcolor{blue}{its} facet, say $b$, is adversarial then $x$ is not in the convex hull of the remaining points $a,c,d$, as shown in Figure~\ref{fig:NC}(b). In the following, we show that this example can generalized to arbitrary number of points and arbitrary dimension.


\begin{proposition}
\label{prop:center_nec}
For a set of $N_i$ nodes in general positions, if  $F_i\ge \left\lceil \frac{N_i}{d+1} \right\rceil $, then there exist general examples in which an $F_i$-safe point does not exist.
\end{proposition}
\emph{Proof:} 
Let $a_1,a_2,\ldots,a_{d+1}$ be points that lie on the vertices of a non-degenerate $d$-simplex in $\mathbb{R}^d$ where one of the points is adversarial. A point in the interior of simplex can not be $1$-safe because it does not lie in the convex hull of normal points for an arbitrary choice of adversarial point. Consider a point $x$ that lies on a facet of $d$-simplex defined by, say, points $a_1,a_2,\ldots,a_{d}$. If one of the points, say $a_1$, on the vertices of such a facet is adversarial, then point $x$ can not lie on $(d-1)$-simplex or the convex hull of the remaining points, $a_2,a_3,\ldots,a_{d+1}$. Thus, such a point can not be a $1$-safe point.

Now, if we replace each point on a vertex of $d$-simplex by $\frac{N_i}{d+1}$ points (for simplicity, we assume $N_i$ to be a multiple of $d+1$), then we get a total of $N_i$ points that are clustered on $d+1$ vertices. {Any point $x$ in the interior of simplex} can not be  $\frac{N_i}{d+1}$-safe because if all $\frac{N_i}{d+1}$ points on an arbitrary vertex of the simplex are adversarial, then the convex hull of the remaining points does not contain {the point $x$}. Let us say we choose a point $x$ on a facet of the  $d$-simplex. If $\frac{N_i}{d+1}$ points lying on a vertex of this facet are adversarial, then the convex hull of the remaining points does not contain $x$. Thus, no point $x$ can be an $\frac{N_i}{d+1}$-safe point for this configuration of points. It is easy to see that argument still works if instead of placing all $\frac{N_i}{d+1}$ points on a vertex of $d$-simplex, we place them arbitrary close to the vertex. Thus, the statement holds for points in general positions as well. $\qed$

We note that our proof for the necessary condition for the existence of an $F_i$-safe point is based on the proof of the Centerpoint Theorem's sharpness. Similar constructions of extreme examples of pointsets are well-studied in the literature \cite{mustafa2015k,Neumann1945OnAI}.  


\subsection{Sufficient Condition for $F_i$-Safe Point in Arbitrary Dimension}

In the previous section, we showed that if there are \textcolor{blue}{at least} $\lceil \frac{N_i}{d+1} \rceil$ adversarial nodes, then an $F_i$-safe point \textcolor{blue}{does not exist. In this section, we show that for \emph{any} $d$, if the number of adversarial nodes $F_i$ is less than $\lfloor\frac{N_i}{d+1}\rfloor$, then an $F_i$-safe point always exists. The main idea will be to show that for a set $\mathcal{S}$ of $N_i$ points in $\mathbb{R}^d$, there exists an interior point of depth $\lfloor\frac{N_i}{d+1}+1\rfloor$. In other words, there exists an interior point in the intersection of all convex sets containing at least $\lfloor\frac{N_i}{d+1}+1\rfloor$ points of $\mathcal{S}$. For this, we first prove some intermediate results.}




\begin{lemma}
\label{lem:halfspaces}
Let $\calS$ be a set of $N$ points in $\mathbb{R}^d$ in general positions, where $d\ge 1$. For a set of $d+1$ convex sets each containing at least $\lceil \frac{Nd}{d+1}+1 \rceil$ points from $\calS$, there are at least $d+1$ points of $\calS$ contained in the intersection of all halfspaces. 
\end{lemma}

\emph{Proof: } Let $X_1,X_2,\ldots,X_{d+1}$ be the convex sets, each containing at least $k$ points of $\calS$. We will refer to the number of points of $\calS$ contained in a set $X_i$ as the \textit{size} of set $X_i$ and denote it as $|X_i|$. Now, for the first set  $X_1$ we have,
	$$
	|X_1| \ge k.
	$$
	Since there are only $N-k$ points left, the set $X_2$ must contain at least $k-(N-k)$ points from $X_1$. Thus,
	$$
	|X_1\cap X_2| \ge 2k-N.
	$$
	Similarly, $X_3$ must contain at least $k-(N-|X_1\cap X_2|)$ points from $X_1\cap X_2$, that is,
	$$
	|X_1\cap X_2\cap X_3| \ge 3k-2N.
	$$
	In general, we have, 
	$$
	|X_1\cap X_2\cap \ldots X_i| \ge k-(N-|X_1\cap X_2\cap \ldots X_{i-1}|).
	$$
	It follows that,
	$$
	|X_1\cap X_2\cap \ldots X_{d+1}| \ge (d+1)k-dN.
	$$
	Substituting the value of $k$, we get the following,
	$$
	|X_1\cap X_2\cap \ldots \cap X_{d+1}| \ge (d+1)\left\lceil \frac{Nd}{d+1}+1 \right\rceil-dN.
	$$
	It follows that
	$$
	|X_1\cap X_2\cap \ldots X_{d+1}| \ge d+1,
	$$
	which is the desired claim. $\qed$

The above lemma implies that any $d+1$ convex sets, each of which contains \textit{enough} points from a set $\calS$, must contain at least $d+1$ points from the same set. Now, if the original pointset $\calS$ is in general position, then so must be the $d+1$ points contained in all convex sets. Furthermore, every $d+1$ points in general positions span a $d$-simplex that contains a ball of radius $\epsilon$ in its interior for sufficiently small value of $\epsilon>0$. Thus, we get the following:

\begin{corollary}
\label{cor:gen_suff}
	Let $\calS$ be a set of $N$ points in $\mathbb{R}^d$ in general positions, where $d\ge 1$. For a set of $d+1$ convex sets each containing at least $\left\lceil \frac{Nd}{d+1}+1 \right\rceil$ points from $\calS$, there exists $\epsilon>0$ such that the intersection of all convex sets contains a ball of radius $\epsilon$. 
\end{corollary} 
Next, we state the following version of \emph{Helly's Theorem} \cite{klee1953critical} that will be used in the proof of Theorem \ref{thm:general_suff}.

\begin{theorem}\cite{klee1953critical}
	\label{theorem:ihelly}
	If $\mathcal{K}$ is a finite family of convex sets (or an infinite family of compact convex sets) in $\mathbb{R}^d$, $C$ is a convex set such that for every $d+1$ members of $\mathcal{K}$, there exists a translate\footnote{For a set $C \subset \mathbb{R}^d$, a translate of $C$ is the set $\{c+x: c\in C\}$ for some vector $x\in \mathbb{R}^d$.} of $C$ contained in their intersection, then there is a suitable translate of $C$ contained in the intersection of all sets in $\mathcal{K}$.
\end{theorem}

Now we state a crucial result to prove a sufficient condition for an $F_i$-safe point in arbitrary $d$.
\begin{theorem}
\label{thm:general_suff}
For any set $\calS$ of $N$ points in $\mathbb{R}^d$ in general positions, where $d \ge 1 $, there exists an interior point in the intersection of all convex sets containing at least $\left\lceil \frac{Nd}{d+1}+1 \right\rceil$ points of $\calS$.
\end{theorem}

\emph{Proof:} From Lemma~\ref{lem:halfspaces} and Corollary~\ref{cor:gen_suff}, we know that every $d+1$ convex sets contain an $\epsilon$ radius ball for some $\epsilon>0$. Let $\epsilon_0$ be the minimum over all such $\epsilon$ for all choices of $d+1$ convex sets.\footnote{In general, there are infinitely many such sets but we observe that any such set must contain convex hull of some $d+1$ points. Thus, it suffices to only consider $n\choose d+1$ convex sets, i.e., the convex hulls of any $d+1$ points.} Let $C$ be a ball of radius $\epsilon_0$, and let $\mathcal{K}$ be the family of all convex sets that contain points of $\calS$. Then every $d+1$ members of $\mathcal{K}$ contain a translate of $C$. It follows from Theorem~\ref{theorem:ihelly} that a suitable translate of $C$ is contained in the intersection of all sets in $\mathcal{K}$. Since $C$ is a ball of non-zero radius, it has an interior. It follows that intersection of every convex set that contains points from $\calS$ contains a ball with an interior. $\qed$

\textcolor{blue}{Next, we present the main result in this subsection, that is, a sufficient condition for the existence of an $F_i$-safe point in an arbitrary dimension $d$.}

\begin{theorem}
\label{thm:main_suff_con}
\textcolor{blue}{
For a set of $N_i$ nodes in general positions in $\mathbb{R}^d$, if  the number of adversarial nodes is less than $ \lfloor \frac{N_i}{d+1} \rfloor $, then there always exists  an $F_i$-safe point.
}
\end{theorem}

\textcolor{blue} {
\emph{Proof:} From the assumption of Theorem~\ref{thm:main_suff_con}, we have a set of $N_i$ points in general position of which there are less than  $\lfloor \frac{N_i}{d+1} \rfloor $ adversarial. If we take the convex hull of the remaining points, such of convex hull will contain at least $N_i - \left( \lfloor \frac{N_i}{d+1} \rfloor -1 \right) $ normal points. A family $\calK$ of convex sets each of which contains at least $\left\lceil \frac{N_id}{d+1}+1 \right\rceil$ points from a given set of $N_i$ must include the one that contains the convex hull of normal points. Theorem~\ref{thm:general_suff} states that there is a point in the interior of the intersection of family $\calK$. Therefore, existence of an interior point in the convex hull normal points, i.e., an $F_i$-safe point, is implied. This completes the proof. $\qed$
}

\subsection{Improvement in the Resilience of ADRC Consensus Algorithm}
If a normal agent $i$ uses approximate Tverberg partition~\cite{mulzer2013approximating} to compute an $F_i$-safe point, as is done in \cite{park2017fault}, then the algorithm is resilient to $F_i\le\frac{N_i}{2^d}-1$ adversaries in $\mathcal{N}_i$ as compared to the best possible (theoretical) bound \textcolor{blue}{ $F_i \le \frac{N_i-1}{d+1}$ }.
Here, we discuss the usefulness of centerpoint in computing an $F_i$-safe point to improve the ADRC algorithm's resilience.  The main point is that we can compute a centerpoint exactly in two and three dimensions and can compute it with better approximation (compared to Tverberg partition) in higher dimensions using known algorithms. Next, we state the resilience of the centerpoint-based computation of an $F_i$-safe point.

\vspace{-0.3in}
\textcolor{blue}{
\begin{proposition}
\label{prop:CP_practical}
Given a set of $N_i$ points in general positions in $\mathbb{R}^d$, of which $F_i < \mathcal{F}_{max}$ are adversarial, then it is possible to compute an $F_i$-safe point using centerpoint if
\begin{equation}
\label{eq:practical_CP}
\begin{split}
 \mathcal{F}_{max} & = \left\lfloor \frac{N_i}{d+1} \right\rfloor , \hspace{0.2in} \text{for } d = 2,3, \text{ and}\\
\mathcal{F}_{max} & =   \Omega\left(\frac{N_i}{d^2}\right) \hspace{0.7in} \text{for } d > 3.
\end{split}
\end{equation}
Moreover, such an $F_i$-safe point can be computed in $O(N)$ and $O(N^2)$ times in $d=2$ and $3$, respectively, and in $O\left( N^{c\log d} (2d)^d \right)$ time in $d>3$ dimensions.
\end{proposition}
}

\textcolor{blue}{\emph{Proof:} By Theorem \ref{thm:safe_center}, we know that an $F_i$-safe point is an interior centerpoint of $N_i$ points, where $F_i\le \frac{N_i}{d+1}-1 $. An interior centerpoint can be computed by taking the centroid of $d+1$ distinct centerpoints whenever an interior centerpoint exists. Therefore, we only need algorithms to compute centerpoints. If $d=2$, a centerpoint can be computed in linear time using the algorithm presented in \cite{jadhav1994computing}. If $d=3$, a centerpoint can be computed in $O(N_i^2)$ time by running Chan's algorithm \cite{chan2004optimal} four times and then computing the centroid of the four centerpoints returned. In higher dimensions $(d\ge4)$, the complexity of finding a centerpoint in the general dimension is unknown. However, Miller and Sheehy provided an algorithm in \cite{miller2010approximate} to compute an approximate centerpoint of depth $\Omega\left(\frac{N}{d^2}\right)$ in $O(N_i^{c\log d}(2d)^d)$ time. As a result, we can use the algorithm to get an $F_i$-safe point with $F_i < \mathcal{F}_{max}= \Omega\left(\frac{N}{d^2}\right)$. $\qed$}

\textcolor{blue}{\textbf{Remark 1}} By comparing \eqref{eq:practical} and \eqref{eq:practical_CP}, observe that centerpoint-based computation of $F_i$-safe point improves the resilience of ADRC algorithm as compared to the Tverberg partition-based computation. Using centerpoint, we achieve optimal resilience in two and three dimensions, whereas in higher dimensions, the resilience improves from $\Omega({N_i/2^d})$ (based on Tverberg partition) to $\Omega(N_i/d^2)$ adversaries in the neighborhood of a normal agent $i$.

\textcolor{blue}{\textbf{Remark 2} The point returned by the algorithm by Miller and Sheehy \cite{miller2010approximate} has a centerpoint depth of $\Omega\left(\frac{N_i}{d^{r/{r-1}}}\right)$ for any positive integer $r > 1$. For instance, if we consider $r=2$, then we get a point of depth $\Omega(N_i/d^2)$. By increasing $r$, the quality of approximation, and hence the bound on the number of adversarial agents improves and approaches $\Omega(N_i/d)$. However, it comes at the cost of increasing time complexity as the runtime of the algorithm is $O\left( N_i^{c\log d} (rd)^d \right)$ for an integer $r>1$ \cite{miller2010approximate}. Moreover, in another work, Clarkson et al. has proposed an algorithm in \cite{clarkson1996approximating} that computes a point of depth $\Omega(N_i/d^2)$ with high probability in time $O(\log^2 d \log\log N_i)$.}

{\color{blue}{
\subsection{Comparison of the Centerpoint-based Computation of an $F_i$-Safe Point with Other Approaches}
\label{sec:new_compare}

In \cite[Sectoin 6.2]{mendes2015multidimensional}, a linear program (LP) is presented to compute an $F_i$-safe point, where $F_i<N_i/(d+1)$. The proposed LP has a total of {\scriptsize{$\dbinom{N_i}{N_i-F_i} (d+1-N_i-F_i)$}} constraints in {\scriptsize{$d+\dbinom{N_i}{N_i-F_i}$}} variables. Hence, their algorithm's time complexity is at least exponential in $F_i$. Since adversarial agents can be as many as a fraction of the total agents, the LP complexity in \cite{mendes2015multidimensional} is exponential in $N_i$. Consequently, as $F_i$ grows with $N_i$, their algorithm's time complexity is too high to be practical even for agents in a two-dimensional space. We note that a centerpoint can also be computed exactly (thus, achieving the same resilience in $O(N_i^{d-1})$ time for $d\ge 3$ using a better LP method proposed in \cite{chan2004optimal}. This is a significant improvement over the LP in \cite{mendes2015multidimensional} because the time complexity is exponential in $d$ instead of the number of agents. Nevertheless, even the algorithm in \cite{chan2004optimal} becomes impractical in higher dimensions. Therefore, we propose that an approximate centerpoint be computed in higher dimensions using \cite{miller2010approximate}, resulting in the resilience of $\Omega(N_i/d^2)$.   

In \cite{WangNLA2019}, a quadratic program is presented to solve the resilient convex combination problem to compute an $F_i$-safe point. The complexity of the algorithm (\cite[Section 3.2]{WangNLA2019}) is $O(d (N_i r)^3)$, which is linear in $d$. Here, the parameter $r$ is defined (in Section 3.1 of their paper) as {\scriptsize{$r = \dbinom{N_i - \sigma}{N_i - \sigma - F_i}$}}. Assuming $\sigma = 0$,\footnote{\textcolor{blue}{This is so because in our setting, a normal agent is unaware of the identities of the normal and adversarial agents in its neighborhood.}} the complexity of computing an $F_i$-safe point using the quadratic program is $O\left(d ( {N_i}^{ N_i+1})^3\right)$, where  a fraction of the agents are allowed to be adversarial. As a result, the time complexity of the quadratic program in \cite{WangNLA2019} is worse than the LP proposed in \cite{mendes2015multidimensional}. Similarly, in \cite[Section IV]{yan2020resilient} authors present an algorithm to compute an $F_i$-safe point (that they call a ``middle point'') in time $O((pr)^3)$, where $p=dF_i+1$, and $r = {{(d+1)F_i+1}\choose{F_i}}$. Note that the parameter $r$ is exponential in the number of adversarial agents.

In summary, the centerpoint-based approach provides fundamental limits on the computation of an $F_i$-safe point. An immediate result of this approach is that in general, computation of an $F_i$-safe point, where the maximum value of $F_i$ is $\Omega(N_i/d)$, is a challenging problem. Since checking whether a given point is a centerpoint or not is \emph{co-NP-Complete} \cite{TengPhD1991}, we deduce that checking whether a point is  an $F_i$-safe point or not in general is also \emph{co-NP-Complete}. Further, our centerpoint approach and the proposed algorithms offer improvements in time complexity over the algorithms in \cite{mendes2015multidimensional,yan2020resilient,WangNLA2019}. 
}}
\section{Numerical Evaluation}
\label{sec:Evaluation}
We perform simulations to illustrate resilient consensus in multirobot systems in two dimensions using centerpoint, and compare it with the one using approximate Tverberg partition \cite{park2017fault}.\footnote{Our code is available at \url{https://github.com/JianiLi/MultiRobotsRendezvous}} We model interconnections between robots using the following graphs: 

\begin{itemize}
    \item A disk graph $\mathcal{G}_d = (\mathcal{V}, \mathcal{E}(t))$, in which each node $i\in\mathcal{V}$ (representing a robot) has a sensing radius $r$ and $(j,i)\in\mathcal{E}(t)$ if and only if $||x_j(t) - x_i(t)|| \le r$.  
    \item A fixed {undirected} graph $\mathcal{G} = (\mathcal{V}, \mathcal{E})$ whose edge set does not change over time.
\end{itemize}

At each iteration $t$ of the consensus algorithm, a normal robot $i$ computes an $F_i$-safe point $s_i(t)$ of its neighbors' positions (using centerpoint or approximate Tverberg partition), and calculates its new position using \eqref{eq:ADRC}. In our experiments, we set $\alpha_i(t) = 0.8$, for all $t$.


\subsection{Centerpoint based resilient consensus}
\begin{figure}[htb]
\centering
\begin{subfigure}[b]{0.155\textwidth}
\centering
\includegraphics[scale = 0.14]{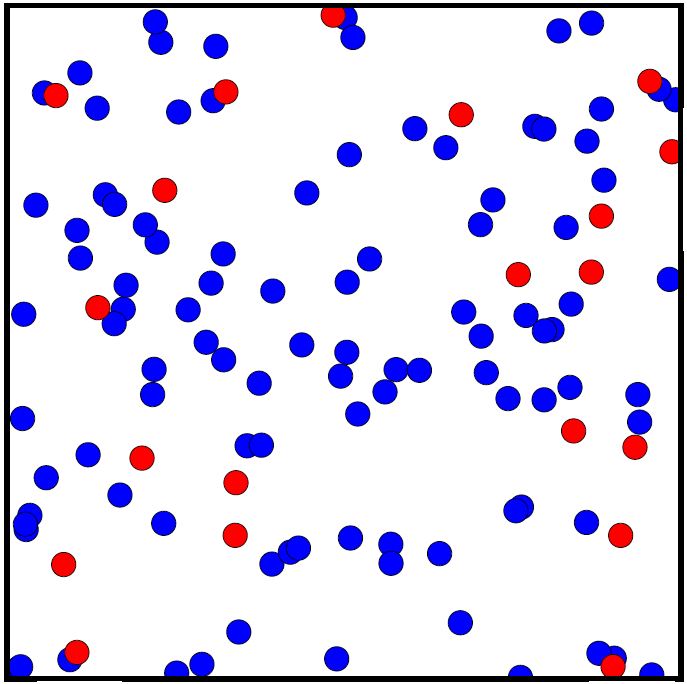}
\caption{}
\end{subfigure}
\begin{subfigure}[b]{0.155\textwidth}
\centering
\includegraphics[scale = 0.14]{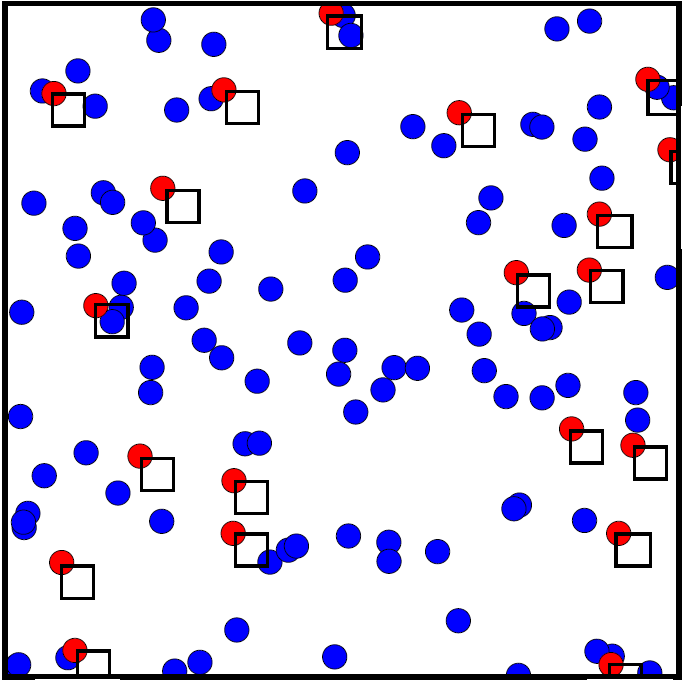}
\caption{}
\end{subfigure}
\begin{subfigure}[b]{0.155\textwidth}
\centering
 \includegraphics[scale = 0.14]{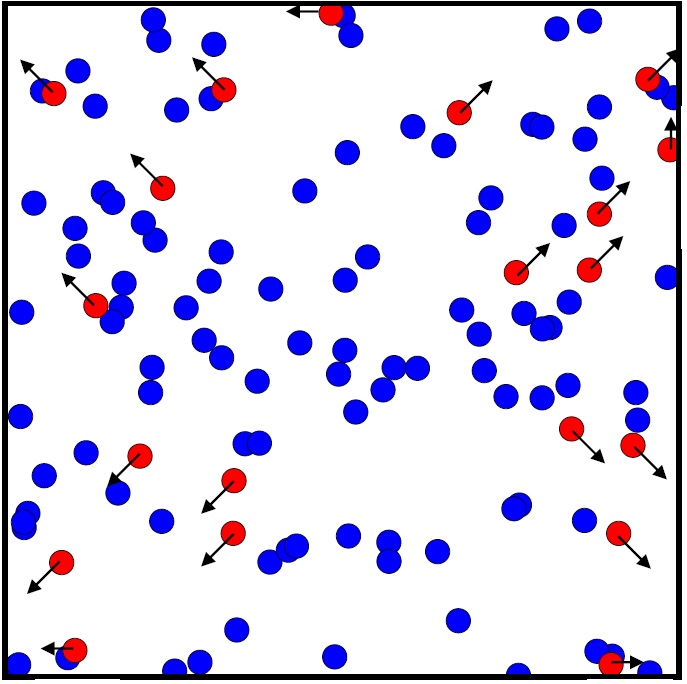}
\caption{}
\end{subfigure}
\caption{Robots' initial positions with (a) stationary, (b) oscillating  and (c) move-away adversaries.}
\label{fig: centerpoint consensus initial topology}
\end{figure}

We simulate a group of 120 robots, {out of which 100 are normal and 20 are adversarial}. They are deployed in a planar region $\mathcal{W} = [-1,1] \times [-1,1] \in \mathbb{R}^2$ as shown in \figref{fig: centerpoint consensus initial topology}(a), where nodes in blue are normal robots executing consensus and nodes in red represent adversarial robots. We consider three types of adversarial behaviors:
\begin{itemize}
    \item \emph{Stationary} -- each red node has a fixed position that does not change throughout the simulation.
    \item \emph{Oscillating} -- each red node changes its position by moving from corner to corner within the square of length 0.1, as depicted in \figref{fig: centerpoint consensus initial topology}(b).
    \item \emph{Move-away} -- each red node moves towards the closest corner of the region $\mathcal{W}$ from its initial position 
    and then stays there, as shown in \figref{fig: centerpoint consensus initial topology}(c).
\end{itemize}

\begin{figure}[htb]
\centering
\begin{subfigure}[b]{0.155\textwidth}
\centering
\includegraphics[scale = 0.14]{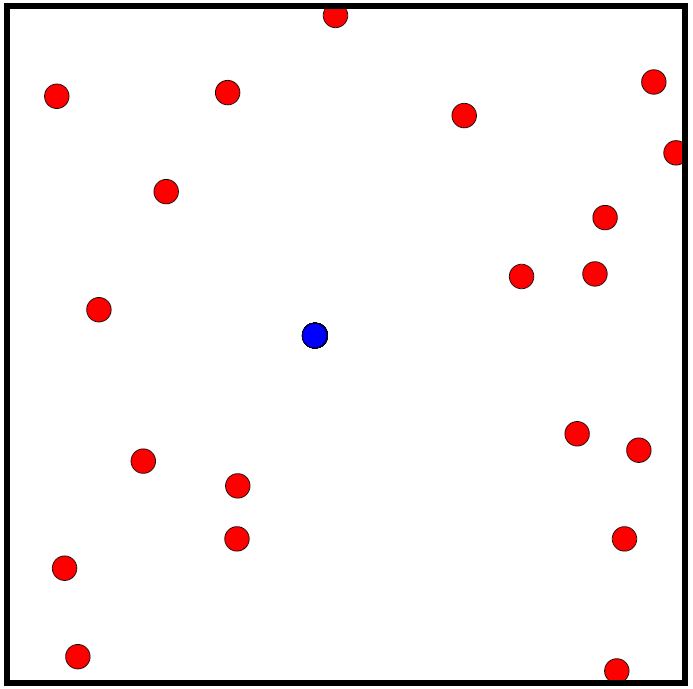}
\caption{}
\end{subfigure}
\begin{subfigure}[b]{0.155\textwidth}
\centering
\includegraphics[scale = 0.14]{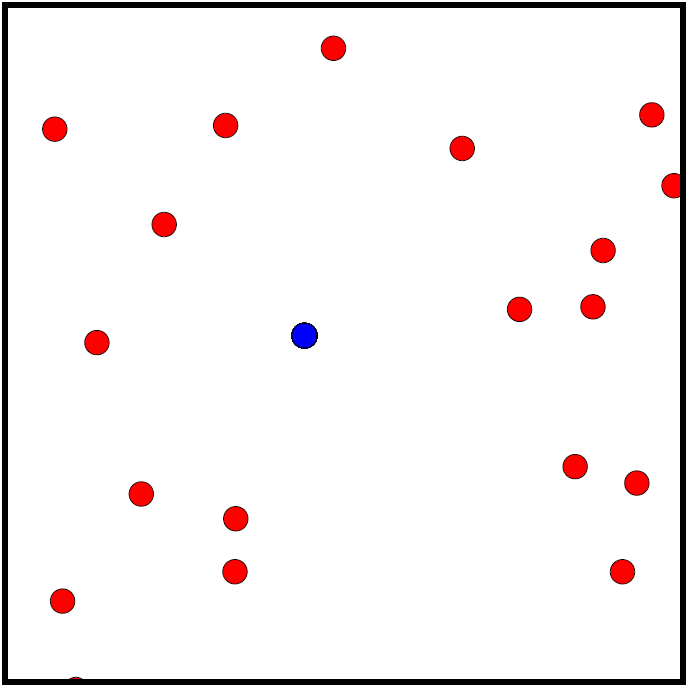}
\caption{}
\end{subfigure}
\begin{subfigure}[b]{0.155\textwidth}
\centering
 \includegraphics[scale = 0.14]{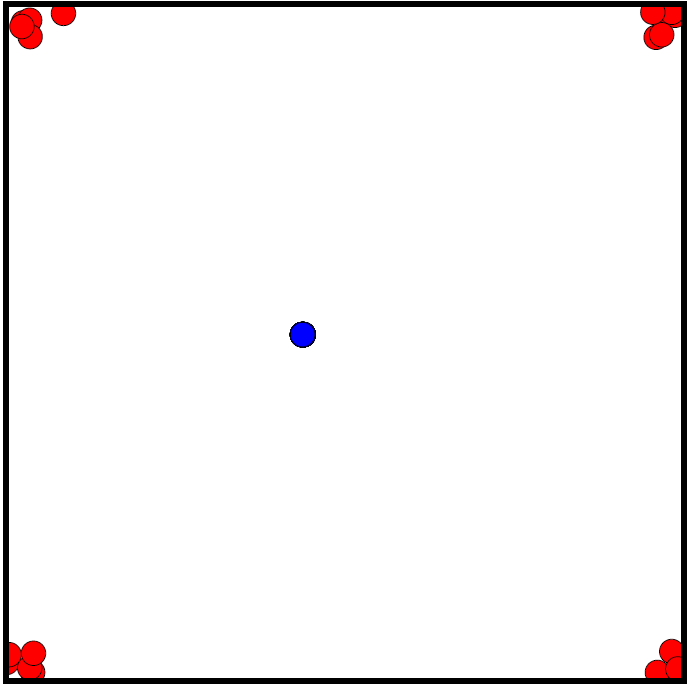}
\caption{}
\end{subfigure}
\caption{Robots' positions after consensus with (a) stationary, (b) oscillating, and (c) move-away adversaries.}
\label{fig: centerpoint consensus evolution}
\end{figure}

\begin{figure}[htb]
\centering
\begin{subfigure}[b]{0.155\textwidth}
\centering
\includegraphics[width=1.1in, height = 1.7in, trim=0.4cm 0.4cm 0.4cm 0.4cm]{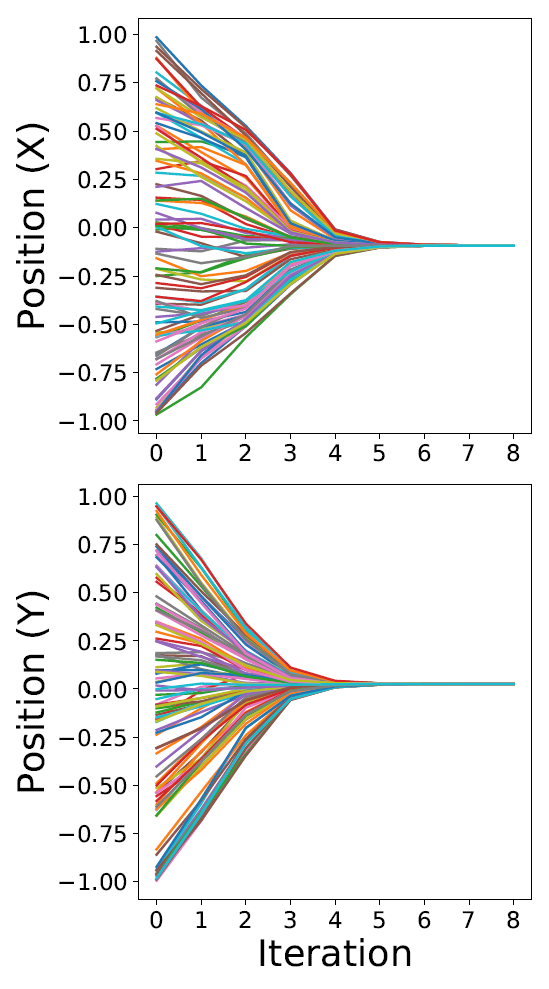}
\caption{}
\end{subfigure}
\begin{subfigure}[b]{0.155\textwidth}
\centering
 \includegraphics[width=1.1in, height = 1.7in, trim=0.4cm 0.4cm 0.4cm 0.4cm]{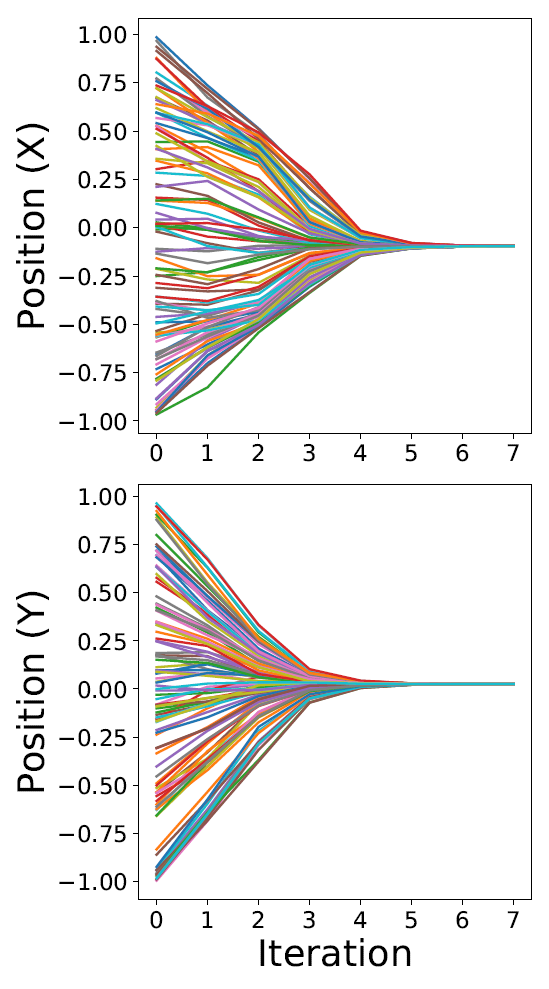}
\caption{}
\end{subfigure}
\begin{subfigure}[b]{0.155\textwidth}
\centering
\includegraphics[width=1.1in, height = 1.7in, trim=0.4cm 0.4cm 0.4cm 0.4cm]{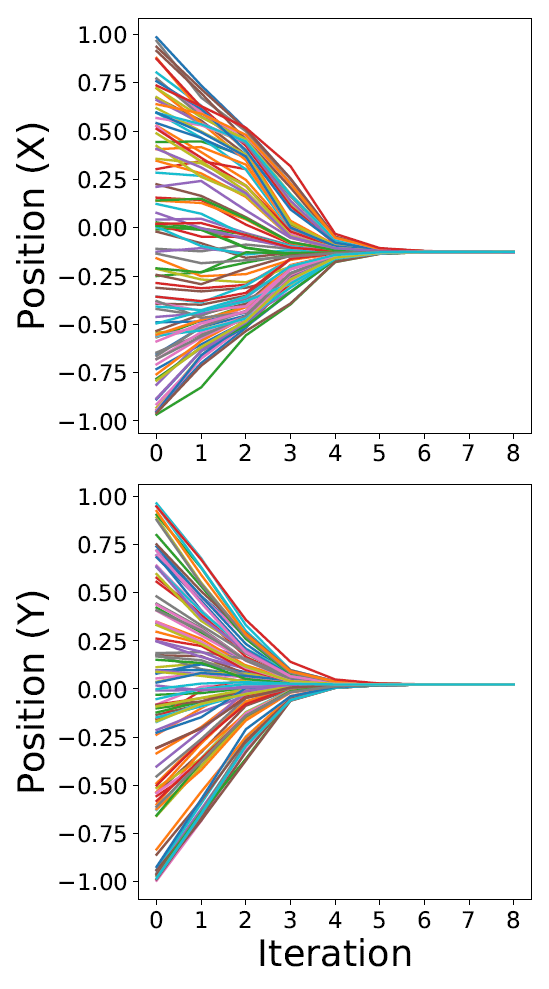}
\caption{}
\end{subfigure}
\caption{Normal robots' positions as functions of iterations in the case of (a) stationary, (b) oscillating, and (c) move-away adversaries.}
\label{fig: centerpoint consensus positiion change}
\end{figure}

The interconnection topology is captured by a disk graph with a fixed sensing radius of $r = 0.45$ for every normal robot. Each normal robot executes consensus algorithm based on centerpoint. Consensus is guaranteed if the number of adversarial robots in the neighborhood of each normal robot $i$ is $F_i \le \left(\left\lfloor \frac{{N}_i}{3}\right\rfloor-1\right)$ \textcolor{blue}{and the sequence of graphs induced by normal agents is repeatedly reachable. Both of these conditions are satisfied in all the simulations here.} Consequently, normal robots achieve consensus in the presence of all three types of adversarial nodes, as shown in Figures \ref{fig: centerpoint consensus evolution} and \ref{fig: centerpoint consensus positiion change}.

\subsection{Comparison of Approximate Tverberg Partition and  Centerpoint-based Resilient Consensus}
\begin{figure}[htb]
\centering
\begin{subfigure}[b]{0.23\textwidth}
\centering
 \centering
        \includegraphics[width=1.75in]{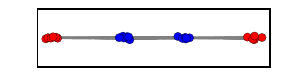}\\
        \includegraphics[width=1.75in]{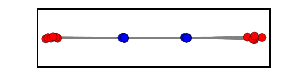}\\
        \includegraphics[width=1.75in]{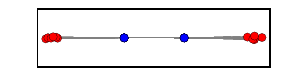}\\
        \caption{}
\end{subfigure}
\begin{subfigure}[b]{0.23\textwidth}
\centering
        \includegraphics[width=1.75in]{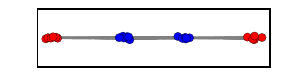}\\
        \includegraphics[width=1.75in]{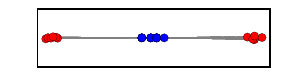}\\
        \includegraphics[width=1.75in]{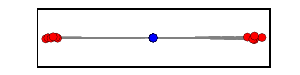}\\
\caption{}
\end{subfigure}
\caption{The evolution of robots' positions (from top to down) in the case of (a) approximate Tverberg partition-based and the (b) centerpoint-based algorithms.}
\label{fig: simple: Robots position evolution}
\end{figure}

To demonstrate the improved resilience of centerpoint-based consensus compared to the Tverberg partition-based one, we consider stationary adversarial nodes and assume that the network graphs are fixed. In our first example, we consider a group of 28 robots deployed in a planar region $\mathcal{W} = [-1.5,1.5] \times [-0.375, 0.375] \in \mathbb{R}^2$ as shown in \figref{fig: simple: Robots position evolution}. There are six adversarial nodes (red) on the left and six on the planar region's right side. At the same time, 16 normal robots (blue) are divided equally into two clusters in the middle part of the region, each containing eight robots. All of the 16 normal robots are pair-wise adjacent to each other. Moreover, each normal robot is adjacent to six adversarial robots located on its side. Figures \ref{fig: simple: Robots position evolution} shows the evolution of robots' positions using approximate Tverberg partition-based and centerpoint-based consensus algorithms. As illustrated, the approximate Tverberg partition-based algorithm fails to make the normal robots converge to one point, but the centerpoint-based algorithm succeeds. The main reason is that each normal robot $i$ has {${N}_i=22$ robots in its neighborhood} (including the robot itself), of which $6$ are adversarial. The approximate Tverberg partition-based algorithm is resilient to $\left( \left\lceil \frac{{N}_i}{4}\right\rceil - 1 \right) = 5$ adversaries, 
whereas the centerpoint based algorithm is resilient to $\left(\left\lfloor \frac{{N}_i}{3}\right\rfloor -1\right) = 6$ adversaries. 

\section{Conclusion}
\label{sec:Con}
 
To achieve resilient vector consensus, normal agents in a network depend on a scheme whereby they compute a state that is a convex combination of their normal neighbors' states. We discussed the merits of using the notion of centerpoint for this purpose. 
For a normal agent $i$, we provided a geometric characterization---using centerpoint---of states that are guaranteed to lie in the convex hull of agent $i$'s normal neighbors' states when the number of adversarial agents is limited to a $1/(d+1)$ fraction of the size of the neighborhood of $i$. It also followed that the upper bound on the number of adversarial agents is best possible in the worst case. We proposed to use well-known efficient algorithms to compute exact centerpoints in two and three dimensions. For higher dimensions, by using the approximate centerpoint algorithm proposed in \cite{miller2010approximate}, we showed that the resilience of vector consensus algorithm to adversarial agents in the neighborhood of a normal agent $i$ could be significantly improved from $\Omega(N_i/2^d)$, which is due to approximate Tverberg partition \cite{park2017fault}, to $\Omega (N_i/d^2)$. 
\textcolor{blue}{The centerpoint-based aggregation of data in the presence of adversarial agents can be used in a various other applications, including resilient distributed diffusion and learning. In the future, we aim to further improve the resilience of vector consensus algorithms by introducing trusted and diverse agents within the network \cite{abbas2017improving,mitra2018impact}.} 

\bibliographystyle{ieeetr}        
\bibliography{References}           
\end{document}